\documentclass[]{article}
\usepackage{amsmath,amssymb,graphicx}
\usepackage{fullpage}

\def\##1{{\underline #1}}
\def\=#1{\underline{\underline #1}}

\def\*#1{{\cal {#1}}}

\def\.{\mbox{ \tiny{$^\bullet$} }}

\def\lambdao{\lambda_{\scriptscriptstyle 0}}

\def\ko{k_{\scriptscriptstyle 0}}

\begin{document}

\begin{center}
{\bf Experimental excitation of multiple surface-plasmon-polariton waves with 2D gratings}\\
\vspace{0.5cm}
Liu Liu,$^1$ {Muhammad Faryad},$^{2,3}$ {A. Shoji Hall},$^{4,5}$ {Sema Erten},$^2$ {Greg D. Barber},$^4$
{Thomas E. Mallouk},$^4$ {Akhlesh Lakhtakia},$^2$ and {Theresa S. Mayer}$^1$\\
\vspace{0.5cm}

$^1$Department of Electrical Engineering, Pennsylvania State University, University Park, PA 16802, USA\\
$^2$Department of Engineering Science and Mechanics, Pennsylvania State University, University Park, PA 16802, USA\\
$^3$Department of Physics, Lahore University of Management Sciences, Lahore 54792, Pakistan\\
$^4$Department of Chemistry, Pennsylvania State University, University Park, PA 16802, USA\\
$^5$Department of Chemistry, Massachusetts Institute of Technology, Cambridge, MA 02139, USA\\

\end{center}

\begin{abstract}The excitation of multiple SPP waves as Floquet harmonics was demonstrated in structures fabricated as one-dimensional photonic crystals (PCs) on top of two-dimensional  gold gratings. Each period of the PC comprised nine layers of silicon oxynitrides of different compositions, and each PC had either two or three periods. Absorptances for obliquely incident $p$- and $s$-polarized light were measured in the $500$--$1000$-nm wavelength regime and the sharp bands in the absorptance spectra were compared with the solutions of the underlying canonical boundary-value problem. The excitation of multiple surface-plasmon-polariton (SPP) waves as Floquet harmonics was confirmed. The structures demonstrated broadband  absorption with overall weak dependences on the incidence angle and the polarization state of the incident light, and has potential application  for harvesting solar energy.
\end{abstract}

Multiple surface-plasmon-polariton (SPP) waves can be guided in a broad spectral regime by the planar interface of a metal and a dielectric material that is periodically nonhomogeneous in the direction normal to the interface, as has been predicted theoretically and demonstrated experimentally~\cite{1,2}. One practical way of exciting multiple SPP waves is to make the interface periodically corrugated~\cite{2,7,8}. Hall \textit{et al.}~\cite{2} deposited a 1D photonic crystal (PC) with each period comprising nine different silicon-oxynitride layers on a 1D gold grating. With the incidence plane the same as the grating plane and the PC containing three periods, experimental evidence of multiple SPP waves excited as Floquet harmonics over the $500$--$1000$-nm regime was unambiguously obtained.

When the 1D grating is replaced by a 2D grating, many more Floquet harmonics exist~\cite{3}, leading to increased possibilities of exciting multiple SPP waves. Accordingly, the SPP waves can propagate in several directions not wholly contained in the incidence plane~\cite{1}. In order to verify that hypothesis with future application to thin-film solar cells~\cite{4} with periodically nonhomogeneous active layers~\cite{5}, we experimentally investigated the excitation of multiple SPP waves in a structure comprising a 1D PC deposited on a 2D metal grating with a square lattice. The experimentally obtained dispersions of the excited SPP waves were found to agree with theoretical predictions from the underlying canonical boundary-value problem, in which the 1D PC and the metal occupy contiguous half spaces~\cite{1}. We report our results in this Letter.

\begin{figure*}[htbp]
\centerline{\includegraphics[width=14cm]{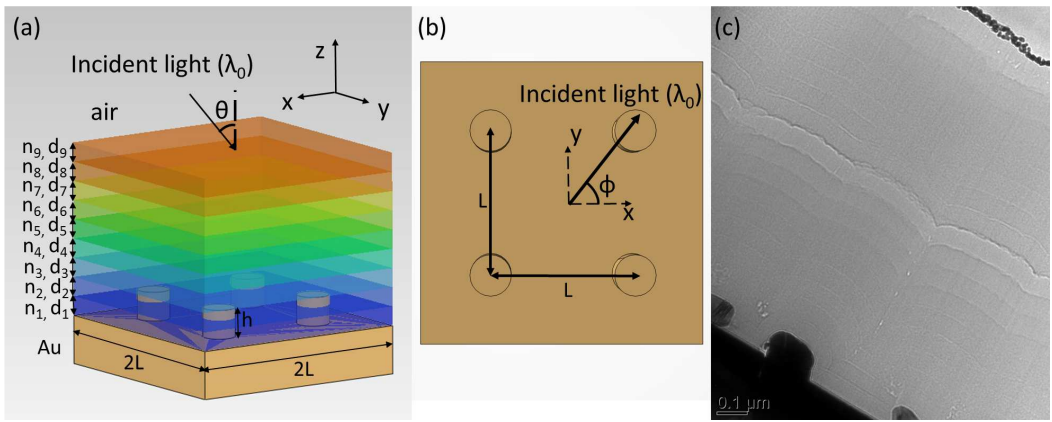}}
\label{geo}
\caption{(a) Schematic of the 1D PC deposited on the 2D metal grating. Only one period of the PC is shown. The wavevector of the incident light is inclined at $\theta$ with respect to the $z$ axis, and (b) at $\phi$ with respect to the 
$x$ axis in the $xy$ plane. (c) Transmission-electron microscopy image of the fabricated sample with 2 periods of PC.}
\end{figure*}
\begin{figure}[ht!]
\centerline{\includegraphics[width=11cm]{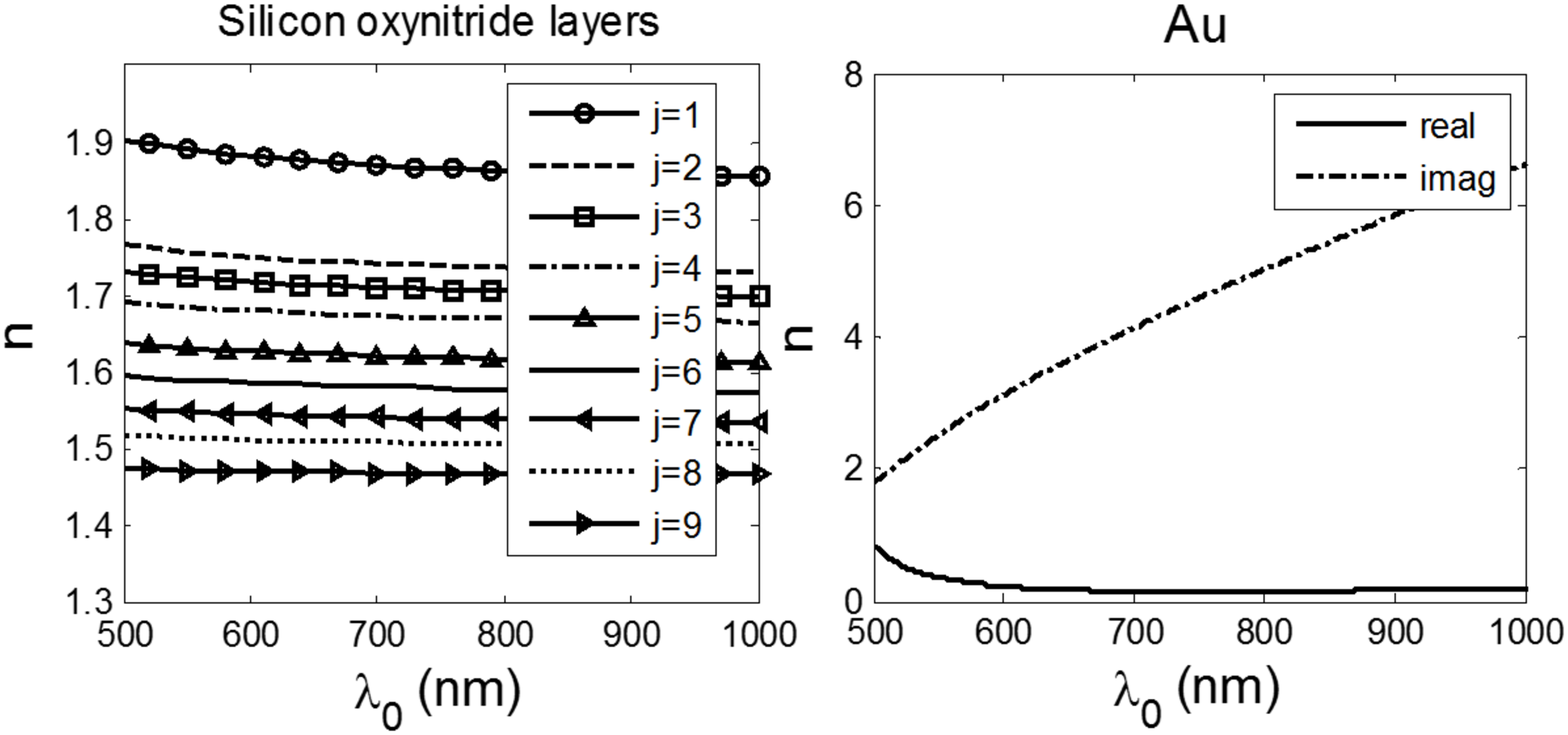}}
\caption{(left) Real refractive indexes of the nine silicon-oxynitride layers in each period of the PC. (right) Real and imaginary parts of the refractive index of Au.
}
\label{index}
\end{figure}

Figures~1(a,b) are schematics of the fabricated structures with two unit cells in both $x$ and $y$ directions. The grating was made of gold (Au) using a template-stripping process described elsewhere~\cite{2}. The lattice constant $L = 350$~nm, the step height is $90$~nm, and the cross-sectional diameter of the step is $190$~nm. The PC on top of the Au grating has either $2$ or $3$ periods (although only $1$ period is shown in the figure). Each period consists of $9$ silicon oxynitride (SiO$_2$)$_a$(Si$_3$N$_4$)$_({}_1{}_-{}_a{}_)$ layers of different compositions identified by $a$ $\in$ $[0,1]$. The refractive index $n_j$ and thickness $d_j$ of the $j$th layer, $j$ $\in$ $[1,9]$, were measured by spectroscopic ellipsometry and from the transmission-electron microscopy image shown in Fig.1(c). The refractive indices are provided in Fig.~\ref{index}, and the thickness of the PC layers are $d_j=[58, 60, 46, 59, 60, 60, 67, 71, 78]$ nm, $j$ $\in$ $[1,9]$.

The total transmittance of the structure is null valued because the metal is thicker than the penetration depth for any free-space wavelength $\lambdao$ $\in$ $[500,1000]$~nm, which was experimentally verified as well. The specular reflectances $R_{s0}$ and $R_{p0}$ were measured for angles of incidence $\theta$ $\in$ $[8,55]$ deg and $\phi = 45$ deg, as shown in Figs.~1(a,b), on a custom spectrometer for incident light of $s$ and $p$ polarization states, respectively. Because all nonspecular reflectances are null valued for the chosen values of $\lambdao$, $\theta$, and $\phi$, the corresponding absorptances of the structure can be calculated as $A_{s,p}=1-R_{s0,p0}$.

\begin{figure*}[ht!]
\centerline{\includegraphics[width=13cm]{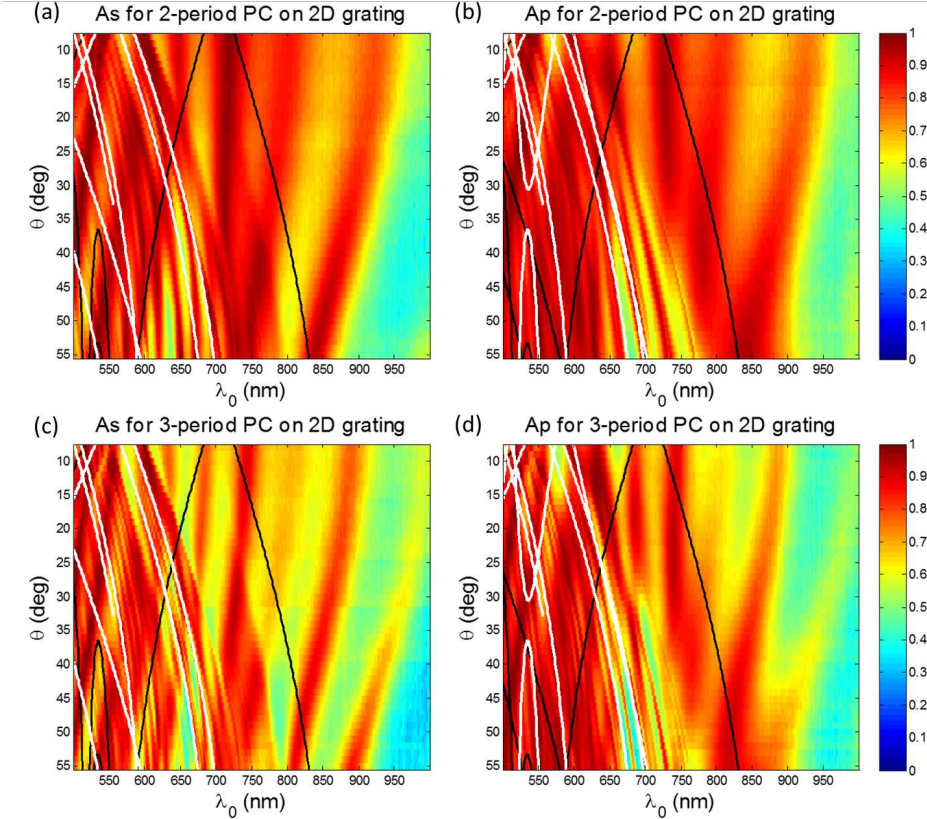}}
\caption{Color maps of the measured absorptances (a,c) $A_s$ and (b,d) $A_p$ of the fabricated structure with the 1D PC of (a,b) $2$ periods or (c,d) $3$ periods, when $\phi=45$~deg. Superimposed curves show predictions from Eq.~(1). The curves that are colored white identify the SPP waves that were actually excited in the structure.}
\label{exp}
\end{figure*}

Figures ~\ref{exp}(a-d) present color maps of the absorptances as functions of $\theta$ and $\lambdao$ for structures with $2$ and $3$ periods of PC. High absorptances are depicted in reddish colors, low absorptances in bluish colors. All four color maps show broadband light absorption over $\theta$ $\in$ $[8,55]$ deg, and we clearly observe reddish ridges of high absorptances, regardless of the polarization state. Each ridge could indicate the excitation of an SPP wave or another optical resonance, such as a waveguide mode~\cite{6}, in the form of a Floquet harmonic.

In order to separate the excitation of the SPP waves from other optical phenomenons, the Floquet harmonics that function as SPP waves must be theoretically identified. For that purpose, we solved the underlying canonical boundary-value problem~\cite{1}, in which the half space $z < 0$ is completely occupied by the 1D PC and the half space $z > 0$
 by Au. This problem yields the dispersion relation of the SPP waves that the metal/PC interface can guide. Given that any SPP wave is a surface wave, the dispersion relations should remain mostly unchanged if the number of periods of the 1D PC is finite but sufficiently large and the metal is sufficiently thick in a practical experiment. 

\begin{figure*}[ht!]
\centerline{\includegraphics[width=13cm]{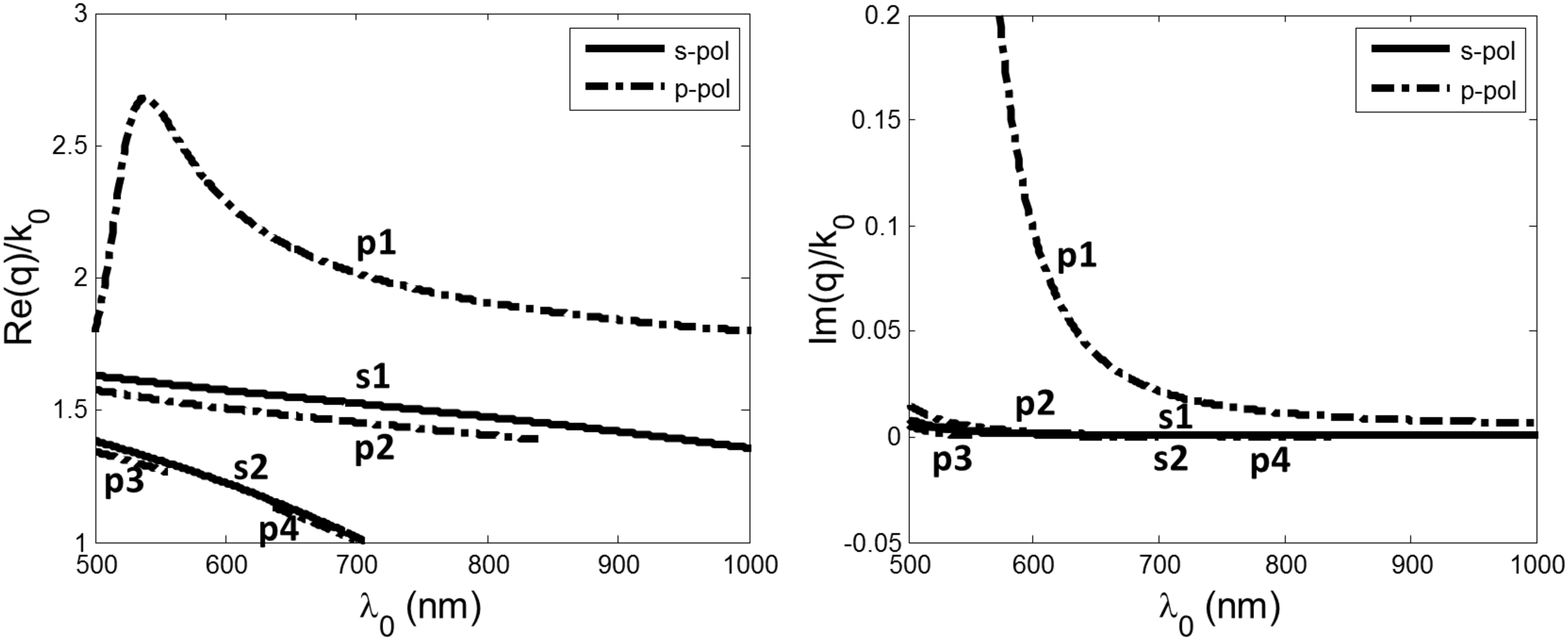}}
\caption{Real (a) and imaginary (b) parts of the normalized wavenumber $q/\ko$ for SPP waves calculated using the canonical boundary-value problem. The wavenumbers are organized into 4 branches of $p$-polarized SPP waves and 2 branches of $s$-polarized SPP waves.}
\label{dis}
\end{figure*}

The wavenumbers $q$ of SPP waves obtained from the dispersion equation for $\lambdao\in[500,1000]$~nm can be organized in six branches, as shown in Fig.~\ref{dis}. These branches are identified as $\left\{p1,p2,p3,p4\right\}$ for $p$-polarized SPP waves, and $\left\{s1,s2\right\}$ for $s$-polarized SPP waves. 

The fields of a Floquet harmonic of order $(m,n)$ vary with $x$ as $\exp[i\ko{x}(\sin\theta \cos\phi+m\lambdao/L)]$ and with $y$ as $\exp[i\ko{y}(\sin\theta \sin\phi+n\lambdao/L)]$, where the free-space wavenumber $\ko=2\pi/\lambdao$,
 $m\in\mathbb{Z}$, $n\in\mathbb{Z}$, and $\mathbb{Z}\equiv\left\{0,\pm1,\pm2,\ldots\right\}$. An SPP wave with 
 wavenumber $q$ is excited as a Floquet harmonic of order $(m,n)$, provided that
\begin{equation}
\pm {\rm Re}(q)=\ko\left\{\left[\sin\theta+(m\cos\phi+n\sin\phi)(\lambdao/L)\right]^2\right.
+\left.\left[(m\sin\phi-n\cos\phi)(\lambdao/L)\right]^2\right\}^\frac{1}{2}\,.
\label{eq1}
\end{equation}
 
\begin{figure}[ht!]
\centerline{\includegraphics[width=11cm]{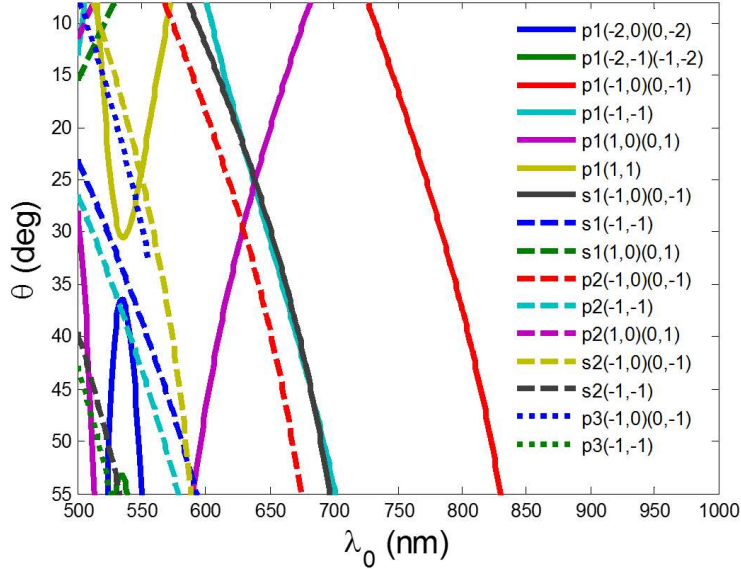}}
\caption{Predictions of $\theta$ as a function of $\lambdao$ using  Eq.~(\ref{eq1}) where an SPP wave could be excited as a Floquet harmonic of order $(m,n)$ when $\phi=45$~deg.}
\label{pred}
\end{figure}

With all  values of $q$ calculated for $\lambdao$ $\in$ $[500,1000]$~nm available in Fig.~\ref{dis}, values of $\theta$ $\in$ $[8,55]$ deg were found for various combinations of $m$ and $n$ to satisfy Eq.~(\ref{eq1}) for $\phi=45$ deg. The 16 curves in Fig~\ref{pred} present those values of $\theta$ as theoretical predictions. Each curve is associated with a specific branch ($p1$-$p4$, $s1$, and $s2$ in Fig.~\ref{dis}) and a specific order $(m,n)$. As $\phi=45$ deg, the wave vectors of the Floquet harmonics of order $(m,m)$, but not of $(m,n \neq m)$, coincide with the projection of the wave vector of the incident plane wave in the $xy$ plane. 

The curves in Fig.~\ref{pred} are also superimposed on the color maps in Fig.~\ref{exp}. Some of those curves coincide with high-absorptance ridges, and are colored white; the other curves are colored black. Each white curve identifies an SPP wave that was experimentally excited as well as the order of the Floquet harmonic that it was excited as. Because each white curve is present whether the number of periods of the PC is two or three, the particular SPP wave is localized to the PC/metal interface on the PC side within three periods at most.  

When the incident plane wave is $p$ polarized, both the $p$-polarized and $s$-polarized SPP waves are excited:  $p1(-2,0)$, $p1(0,-2)$, $p1(-1,-1)$, $p1(-1,1)$, $p1(1,-1)$, $s1(-1,0)$, $s1(0,-1)$, $s1(1,0)$, $s1(0,1)$, $p2(-1,0)$, $p2(0,-1)$, $s2(-1,0)$, $s2(0,-1)$, $p3(-1,0)$, and $p3(0,-1)$. Similarly, when $s$-polarized plane wave is made incident, SPP waves of both linear polarization states are excited: $s1(-1,-1)$, $s1(1,0)$, $s1(0,1)$, $p2(-1,0)$, $p2(0,-1)$, $s2(-1,0)$, $s2(0,-1)$, $s2(-1,-1)$, $p3(-1,0)$, and $p3(0,-1)$. This is because depolarization occurs due to the grating being 2D. Let us also note that not all SPP waves excited are guided by the PC/metal interface to propagate in the incidence plane. 

Each black curve in Figs.~\ref{exp}(a-d) identifies a prediction from Eq.~(1) that could not be observed experimentally. For $s$-polarized incidence, the SPP waves represented by the $p1$ branch could not excited.  For $p$-polarized incidence, the SPP waves represented by $p1(\pm1,0)$ and $p1(0,\pm1)$ could not be observed. This may be due to the fact that the $p1$ branch represents SPP waves that are very tightly bound to the PC/metal interface~\cite{Faryad2012}. Thus surface roughness of the corrugated PC/metal interface might have deviated the excitation of an SPP wave in the $\lambdao\times\theta$ from the predictions of the canonical boundary-value problem based on the planar interface.

In conclusion, we experimentally demonstrated the excitation of multiple SPP waves as Floquet harmonics in the grating-coupled configuration wherein a  2D grating of Au was used. The structures were constructed as either 2- or 3-period 1D PCs on top of 2D gold gratings. Absorptances of $p$- and $s$-polarized light were measured and mapped against wavelength and incidence angle with respect to the thickness direction.  High-absorptance ridges in the maps were compared with theoretical predictions from the underlying canonical boundary-value problem, and the excitation of multiple SPP waves as Floquet harmonics was confirmed.  \\

\vspace{0.3cm}

\noindent{\it Acknowledgments.} This work was supported by the National Science Foundation under Grant No. DMR-1125591. Fabrication experiments were performed at the Pennsylvania State University Materials Research Institute Nanofabrication Laboratory, which is supported by the National Science Foundation under Cooperative Agreement No. ECS-0335765. A.L. is  grateful to Charles Godfrey Binder Endowment for partial support of this work.

\end{document}